\documentclass[]{spie}  

\usepackage{amsmath,amsfonts,amssymb}
\usepackage{graphicx}
\usepackage[colorlinks=true, allcolors=blue]{hyperref}
\usepackage[table]{xcolor}  
\usepackage{booktabs}       
\usepackage{multirow}       
\usepackage{pifont}         
\usepackage{booktabs}
\usepackage{siunitx}
\usepackage{subfig}
\usepackage{cleveref}

\definecolor{lightgreen}{RGB}{144, 238, 144}
\definecolor{lightred}{RGB}{255, 182, 193}
\definecolor{darkgreen}{RGB}{3, 53, 0}
\definecolor{darkred}{RGB}{132, 0, 0}
\definecolor{lightgray}{RGB}{240,240,240}
\definecolor{lightlightgreen}{RGB}{200, 255, 176}

\newcommand{\cmark}{\cellcolor{lightgreen}\textcolor{darkgreen}{\ding{51}}}
\newcommand{\xmark}{\cellcolor{lightred}\textcolor{darkred}{\ding{55}}}

\crefname{figure}{Fig.}{Fig.}
\crefname{table}{Tab.}{Tab.}

\title{A flexible framework for large-scale FDTD simulations: open-source inverse design for 3D nanostructures}

\author[a]{Yannik Mahlau}
\author[a]{Frederik Schubert}
\author[a]{Konrad Bethmann}
\author[b]{Reinhard Caspary}
\author[b]{Antonio Cal\`a Lesina}
\author[a]{Marco Munderloh}
\author[a]{Jörn Ostermann}
\author[a]{Bodo Rosenhahn}
\affil[a]{Leibniz University Hannover, Institute for Information Processing, Appelstraße 9A, 30167 Hannover, Germany}
\affil[b]{Leibniz University Hannover, PhoenixD, Welfengarten 1A, 30167 Hannover, Germany}

\authorinfo{Further author information: (Send correspondence to Yannik Mahlau)\\Yannik Mahlau: E-mail: mahlau@tnt.uni-hannover.de, Telephone: +49 511 762-5319}

\begin{document} 
\maketitle

\begin{abstract}
We introduce an efficient open-source\footnote{The framework is publicly available at \url{https://github.com/ymahlau/fdtdx}} python package for the inverse design of three-dimensional photonic nanostructures using the Finite-Difference Time-Domain (FDTD) method. 
Leveraging a flexible reverse-mode automatic differentiation implementation, our software enables gradient-based optimization over large simulation volumes.
Gradient computation is implemented within the JAX framework and based on the property of time reversibility in Maxwell's equations.
This approach significantly reduces computational time and memory requirements compared to traditional FDTD methods.
Gradient-based optimization facilitates the automatic creation of intricate three-dimensional structures with millions of design parameters, which would be infeasible to design manually.
We demonstrate the scalability of the solver from single to multiple GPUs through several inverse design examples, highlighting its robustness and performance in large-scale photonic simulations.
In addition, the package features an object-oriented and user-friendly API that simplifies the specification of materials, sources, and constraints.
Specifically, it allows for intuitive positioning and sizing of objects in absolute or relative coordinates within the simulation scene.
By rapid specification of the desired design properties and rapid optimization within the given user constraints, this open-source framework aims to accelerate innovation in photonic inverse design.
It yields a powerful and accessible computational tool for researchers, applicable in a wide range of use cases, including but not limited to photonic waveguides, active devices, and photonic integrated circuits. 
\end{abstract}


\section{INTRODUCTION}

Modern three-dimensional design relies heavily on simulation software.
By replacing physical prototypes with digital simulations, engineers can reduce development cycles and costs.
Furthermore, differentiable simulations allow automated design creation using gradient-based optimizations.
In complex scenarios involving thousands or even millions of design parameters, these optimizations often discover designs more efficiently than humans.
However, current electromagnetic simulation and optimization software for the Finite-Difference Time-Domain (FDTD) method falls short of its potential.
The popular open-source software Meep \cite{meep} is limited to CPU hardware, making it too slow for large-scale designs.
In addition, it does not support automatic differentiation.
Another commonly used commercial software, Tidy3D \cite{tidy3d}, offers state-of-the-art performance, but its proprietary per-simulation pricing model creates a significant barrier for the broader research community.
To address these limitations, we developed FDTDX, an open-source electromagnetic simulation and optimization software built for scalability and ease of use.
As it is implemented in the JAX framework, FDTDX provides native GPU support and automatic differentiation capabilities, making it ideal for large-scale 3D design in nanophotonics.

A key challenge in inverse design is efficiently computing gradients to guide the optimization process for a given objective function.
The conventional approach to gradient computation is the adjoint method, which requires users to analytically define adjoint sources during the backward propagation phase.
This analytical derivation process is both time-intensive and susceptible to mathematical errors.
Moreover, it creates an entry barrier for researchers unfamiliar with the adjoint method.
Automatic differentiation offers an alternative approach that computes gradients programmatically, eliminating the need for manual derivations.
Historically, automatic differentiation has been constrained by substantial memory requirements, as it required storing electric and magnetic field values at each time step for every point in the three-dimensional simulation volume.
These memory constraints restricted its application to small-scale simulations.
To overcome these restrictions, FDTDX implements a memory-efficient automatic differentiation approach that leverages the time-reversibility property of Maxwell's equations.
Specifically, the implementation features an inverse update step that transforms the electromagnetic field state from time step t+1 to time step t.

While FDTDX's automatic differentiation capabilities simplify the optimization process, its user-friendly approach extends beyond gradient computation to simulation setup.
Specifying a complex three-dimensional scene programmatically by the absolute position of every object can be tedious and error-prone.
To address this challenge, FDTDX adopts an idea from computer-aided design (CAD) software: the use of relational constraints between objects.
To the best of our knowledge, FDTDX is the first electromagnetic simulation software to incorporate this idea.
FDTDX implements an intuitive API that allows objects to be placed and sized using absolute or relative coordinates, with constraints being resolved automatically.
This high-level interface greatly simplifies the simulation setup, making large-scale FDTD simulations accessible to a wider range of users without sacrificing flexibility.
A feature comparison between FDTDX and other popular FDTD frameworks is shown in \cref{tbl:features}.

We tested our simulation and optimization software in multiple experiments.
Firstly, we performed a scaling analysis to show that FDTDX is capable of running large-scale FDTD simulations.
To this end, we performed simulations at different grid resolutions, which scales the simulations from a few millions up to multiple billions of grid cells.
Then, we optimized a corner element for silicon photonics, which redirects light in by 90 degrees in the small spatial footprint of only $1.6\mu m^2$.
Our optimized corner element achieves a coupling efficiency of 92\% or equivalently an attenuation of -0.36dB.
Lastly, we addressed the challenge of large-scale simulations for fabrication with two-photon polymerization (2PP).
2PP allows for the fabrication of intricate three-dimensional designs, but is restricted to an effective printing field whose size depends on the objective.
Fabricating designs larger than the printing field requires the stitching of multiple fields, leading to alignment errors.
This is particularly evident for printing long waveguides, as errors of just a few micrometers already significantly impede performance.
We optimized a design, which is resilient to random translational offsets offsets of up to $2\mu m$.
An extensive analysis demonstrates the impact of random translational offsets on our device compared to an unmodified waveguide.

\begin{table}[!t]
\centering
\renewcommand{\arraystretch}{1.3}
\caption{Feature comparison of electromagnetic simulation software packages.}
\begin{tabular}{l|c|c|c|c|c|c}
\toprule
\rowcolor{gray!20}
\textbf{Feature} & \textbf{Ceviche}\cite{ceviche} & \textbf{Meep}\cite{meep} & \textbf{Lumerical}\cite{lumerical} & \textbf{OmniSim}\cite{OmniSim} & \textbf{Tidy3D}\cite{tidy3d} & \textbf{FDTDX (ours)} \\
\midrule
3D-Simulation &   \cmark &  \cmark &  \cmark &  \cmark &  \cmark &  \cmark \\
GPU/TPU-capable & \xmark &  \xmark &   \cmark &  \xmark &  \cmark &  \cmark \\
Automatic Differentiation &  \cmark &   \xmark &   \xmark &  \xmark &   \cmark &   \cmark \\
Open-source &  \cmark &   \cmark &  \xmark &   \xmark &   \xmark &   \cmark \\
Relative Positioning/Sizing &   \xmark &  \xmark &  \xmark &  \xmark &  \xmark &  \cmark \\
\bottomrule
\end{tabular}
\label{tbl:features}
\end{table}

In summary, our main \textbf{contributions} are
\begin{itemize}
    \item 
    a FDTD framework capable of rapid simulations named FDTDX. 
    FDTDX offers a high-performance FDTD implementation built on the JAX framework, enabling automatic compilation, computational graph optimization, and efficient GPU execution with built-in multi-GPU scaling capabilities.
    \item
    By releasing FDTDX as open-source software at \url{https://github.com/ymahlau/fdtdx}, we provide the research community with a powerful tool that democratizes access to large-scale electromagnetic design.
    \item 
    The framework implements a memory efficient automatic gradient computation based on the time-reversibility of Maxwell's equations.
    This allows for the large-scale inverse design of three-dimensional nanostructures.
    \item 
    Within our framework, we developed a user-friendly interface for absolute and relative placement and sizing of objects within the simulation volume.
    This intuitive design allows simulations to be performed with little experience in programming.
    \item
    Using our framework, we optimized a corner element for two-dimensional silicon photonics and a stitching element for three-dimensional polymer photonic integrated circuits.
\end{itemize}
Through this open platform, the research community can now collectively advance the frontiers of nanophotonic design.

\section{METHODS}

Our framework implements the FDTD method, which we briefly introduce in the following section.
In addition, we describe different variants for automatic differentiation through an FDTD simulation.
This comparison highlights the advantages of our implementation based on time reversibility.
Lastly, we describe the relational object interface which allows the specification of positional and sizing constraints between objects in the simulation scene.

\subsection{Finite-Differences Time-Domain (FDTD) method}
The FDTD method is based on Maxwell's equations 
\begin{align}
    \frac{\partial H}{\partial t} &= - \frac{1}{\mu} \nabla \times E  \label{eq:faraday}\\
    \frac{\partial E}{\partial t} &= \frac{1}{\epsilon} \nabla \times H, \label{eq:ampere}
\end{align}
where $E$ is the electric and $H$ the magnetic field.
The equations are discretized in time and space according to the Yee grid \cite{kaneyeeNumericalSolutionInitial1966}.
This special grid structure defines the directional components of the electric and magnetic field at interleaved points in time and space such that the curl operator $\nabla$ can be efficiently computed.
The curl updates in Eqs.~\ref{eq:faraday} and \ref{eq:ampere} are used in every time step to update the electric and magnetic field.
FDTDX performs these updates for a given number of time steps and returns the final fields to the user.
It is also possible to access field values from intermediate time steps by using a detector object.

Due to the Yee grid structure, it is necessary to interpolate the fields in time and space if composite metrics using multiple field components are computed.
Instances of such metrics include energy or Poynting flux.
Interpolating the fields at every time step at every grid point can be expensive to compute.
Therefore, FDTDX disables interpolation by default, as approximate measures often suffice, especially during the iterative optimization process of the inverse design.
If the user wants to evaluate a design with high accuracy, interpolation can be turned on.
In the following sections, we disable interpolation for our experiments during optimization and afterwards create evaluation figures using the exact interpolation.

To prevent reflections at the boundaries of the simulation domain, we apply convolutional perfectly matched layers\cite{cpml} (PML) to the boundaries of the simulation volume.
Sources are implemented via the Total-Field Scattered-Field definition \cite{taflove}, which divides the simulation domain into two regions.
The total field region contains both the incident field of the source and the scattered field produced by objects in the simulation scene.
The scattered field region contains only the scattered field, such that incident light is only emitted to a single side of the source.

\subsection{Gradient Computation by Vector Jacobian Product}

FDTD simulations operate as a chain of function calls, where each call updates either the electric or magnetic field.
Applying automatic differentiation to this chain of function calls is challenging as it requires storing all intermediate field values after each update.

One remedy is the recursive halving algorithm \cite{Griewank_2003}, which only saves the fields in logarithmic time steps.
The fields for all other time steps are recalculated using the FDTD forward function from the previous checkpoint.
For a simulation with $n$ time steps, this results in additional $\frac{n}{2}\log_2(n)$ forward function calls.
However, logarithmic memory often still exceeds the capacity of compute clusters for large-scale optimizations.
In this case, a generalization of recursive halving using dynamic programming can be used \cite{dynamic_programming_checkpointing}.
This algorithm, also known as treeverse \cite{treeverse}, employs a divide-and-conquer scheme to automatically calculate the optimal saving strategy for a constant of $k$ checkpoints.
As a result, it is possible to trade memory for runtime depending on the available memory budget by adjusting the number of checkpoints $k$.

One can also exploit the knowledge that Maxwell's equations are reversible in time for linear materials.
Therefore, it is possible to construct the inverse function of the FDTD update, which transforms the electric and magnetic field at time step $t$ to $t-1$.
Instead of saving intermediate values, this function allows us to simply recalculate the previous field values.
Consequently, no intermediate field values have to be saved when only considering Maxwell's equations.
However, the PML at the boundary of the simulation volume are not invertible as they absorb energy, resulting in information loss.
Therefore, the field values at the boundary between the PML and the inner simulation volume need to be saved for every time step.
These intermediate values result in much smaller memory overhead as they are only six two-dimensional slices instead of a full three-dimensional volume.
Additionally, there is only a small runtime increase compared to checkpointing, because the field values do not need to be recalculated over multiple time steps.
This method was first demonstrated by Tang et al. \cite{time_reversible}, but has never been implemented in a powerful automatic differentiation framework.
We follow the previous work of Ref.~\citenum{schubert2024quantized} to implement this algorithm in the JAX-framework.

\subsection{Relational Object Constraints}

For an easy specification of a three-dimensional simulation scene, we introduce an interface for specifying relational object constraints.
In this interface, every object is defined by a cuboid, whose size and position can be specified either directly in absolute coordinates or through relational constraints.
The direct specification can be in real size using metrical units or in integer grid points referring to the discretization of the Yee grid cells.

\begin{figure}[tb]
    \centering
    \subfloat[]{\includegraphics[width=0.45\textwidth]{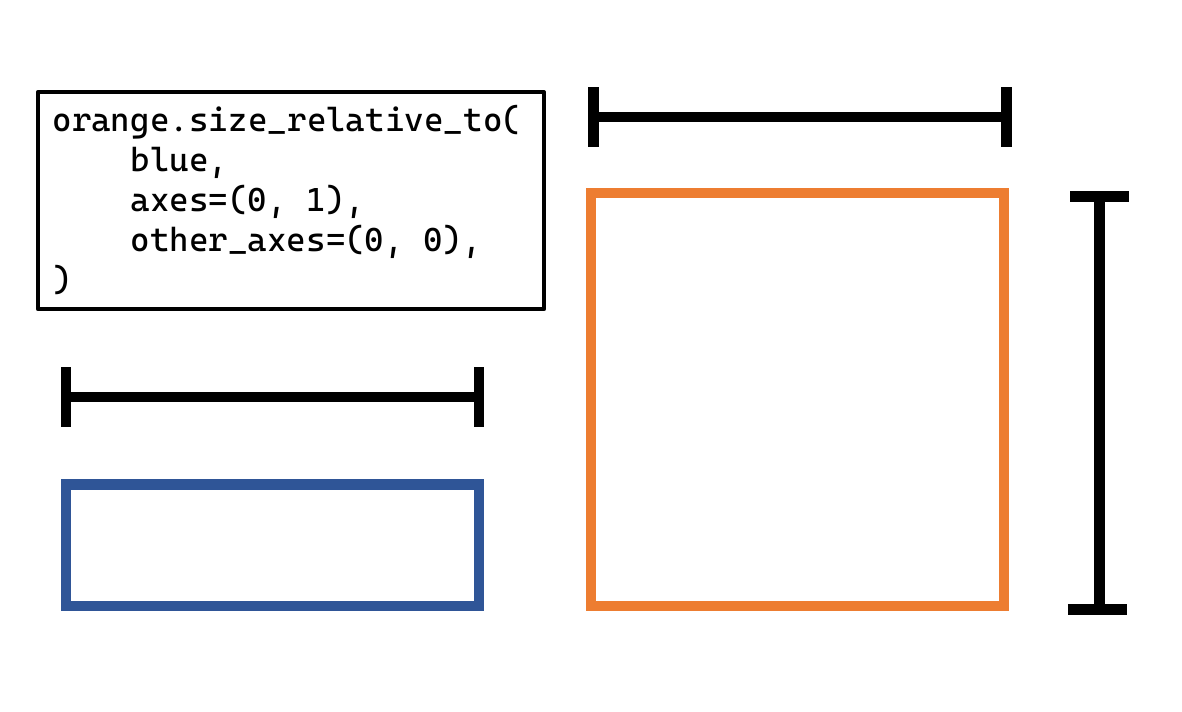}}
    \hspace{0.05\textwidth}
    \subfloat[]{\includegraphics[width=0.45\textwidth]{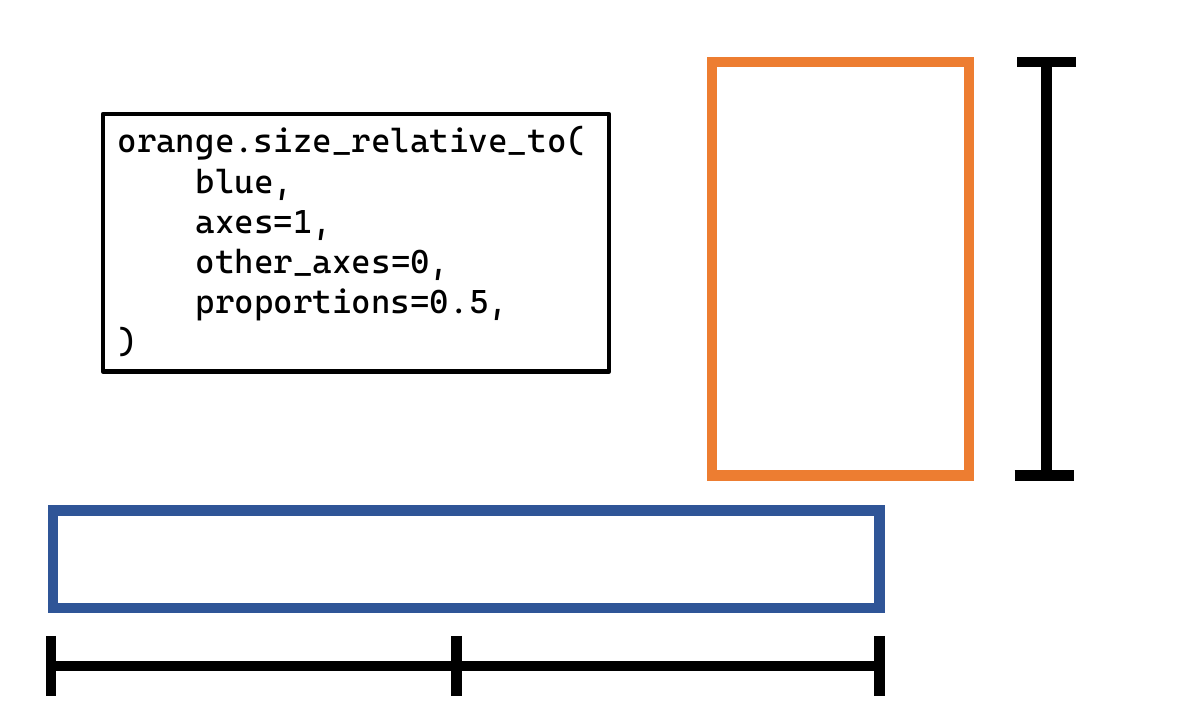}}
    \caption{Relational size constraints across multiple axes and with different proportions. Axis 0 is the x-axis (horizontal) and axis 1 the y-axis (vertical).}
    \label{fig:size_constraints}
\end{figure}

Relational size constraints define the size of one or more axes of the cuboid in relation to the size of one or more axes of another cuboid.
In \cref{fig:size_constraints}, examples of size constraints between two objects are visualized.
It is also possible to specify size constraints such that the size of an object extends to another object in the simulation scene.
The constraints can be specified for the positive and negative directions in an axis.

\begin{figure}[tb]
    \centering
    \subfloat[]{\includegraphics[width=0.3\textwidth]{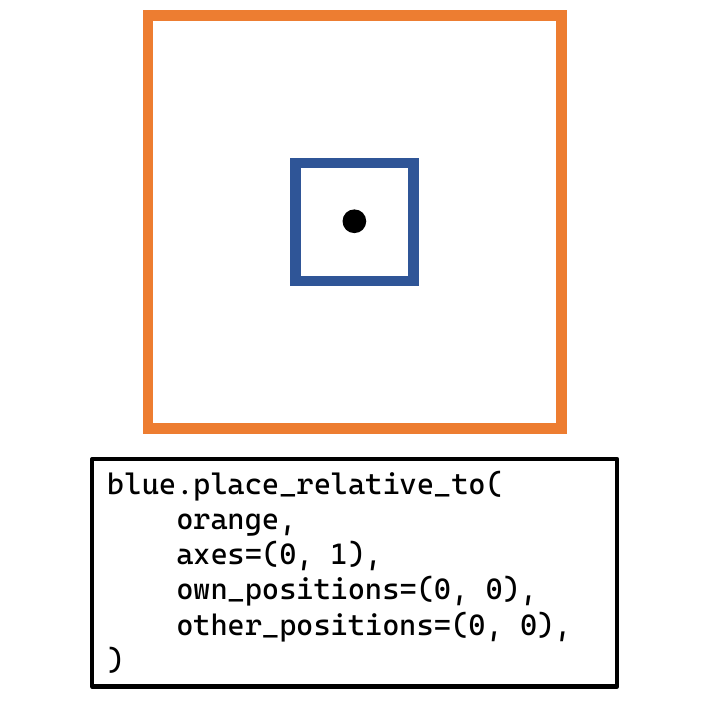}}
    \subfloat[]{\includegraphics[width=0.3\textwidth]{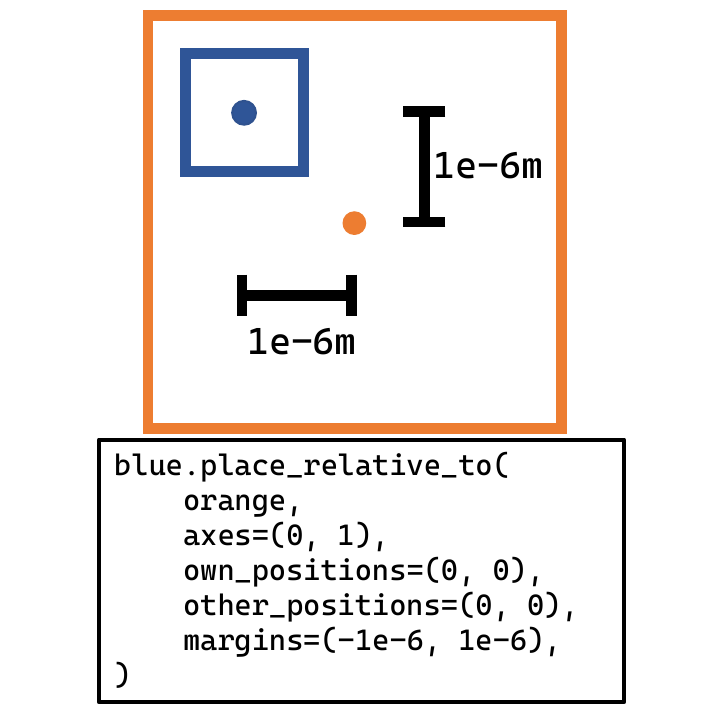}}
    \subfloat[]{\includegraphics[width=0.3\textwidth]{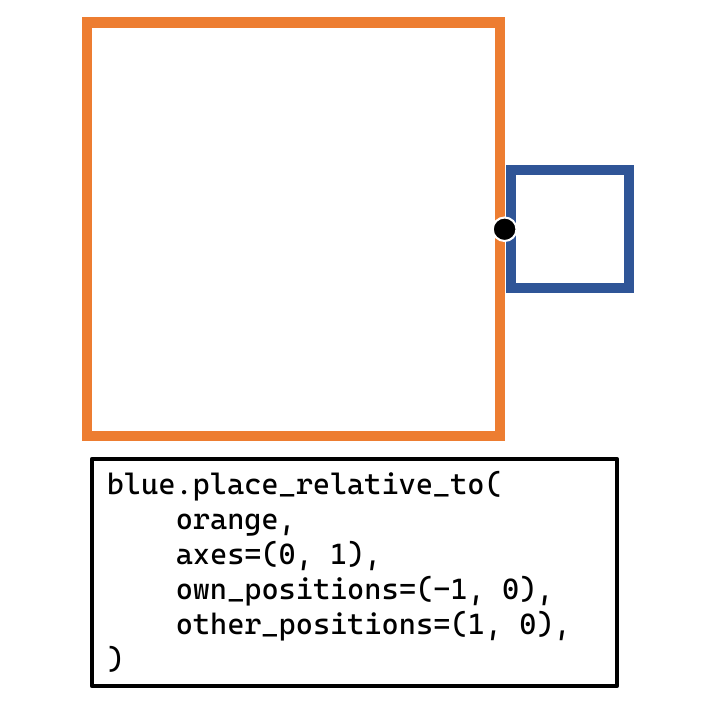}}
    \caption{
    Relational positional constraints across multiple axes and with offsets.
    Axis 0 is the x-axis (horizontal) and axis 1 the y-axis (vertical).
    In (a), the center of the blue rectangle is aligned with the center of the orange rectangle.
    In (b), an offset of $1\mu m$ is applied in negative x and positive y direction.
    In (c), the anchor of the orange object is on the right side (position 1) and the anchor of the blue object on the left side (position -1)
    Therefore, the left side of the blue rectangle is aligned with the right side of the orange rectangle.
    The objects are centered in the y-axis as their anchors also remain centered.
    }
    \label{fig:pos_constraints}
\end{figure}

Positional constraints define the position of a cuboid in one or more axes of the simulation grid.
To this end, an anchor point is defined on both the origin and target cuboid.
By default, the anchor point is placed at the center of the cuboids, which is the origin of the anchor coordinate system.
The anchor coordinate system is normalized, such that a position of $1$ refers to the right side of the cuboid and $-1$ to the left side.
The user can specify the position of the anchor as a vector within this coordinate system.
The constraint system ensures that the anchors of the origin and target cuboid are aligned for the specified axes.
Additionally, it is possible to specify a spatial offset that is applied between the two anchor points.
In \cref{fig:pos_constraints}, examples of positional constraints between two objects are visualized.

The size and position constraints are resolved iteratively until the position and size of all objects are known.
If the size of an object is not specified, it is extended until infinity, i.e. the boundary of the simulation volume.
If there is a conflict between multiple constraints or direct specifications, the algorithms returns an error message to the user.

\section{Results}
We demonstrate the capabilities of our simulation and optimization framework through multiple experiments.
Firstly, we compare the simulation speed of FDTDX with other established simulation frameworks.
Then, we optimize a corner device for silicon photonics, which redirects light between two waveguides placed at a 90-degree angle.
Lastly, we demonstrate that FDTDX is able to optimize intricate three-dimensional designs for 2PP fabrication.
To this end, we optimized a stitching device between two waveguides, which is robust against random translations in the x and y directions.

\subsection{Speed Comparison}

We tested the computational speed of our FDTDX framework compared to other electromagnetic simulation frameworks.
To this end, we simulated the silicon coupling element introduced in Ref.~\citenum{Shen_14_coupler}.
In detail, the simulation volume of this experiment has a size of $6\mu m \times 4 \mu m \times 1.5 \mu m$.
The simulation was run for 200 femtoseconds with a courant factor of 0.99.
We varied the resolution of the Yee grid to analyze the scaling behavior of the different simulation frameworks.
Starting from a resolution of $25nm$ with 2.3 million grid cells, we gradually increased the resolution to $2.5nm$.
At this resolution, the simulation volume consists of 2.3 billion grid cells.
With varying resolution, the number of simulation time steps also changes to satisfy the Courant-Friedrichs-Lewy stability conditions \cite{Courant1928berDP}.

\begin{table}[t]
    \centering
    \sisetup{
        table-format=2.1,
        table-alignment=center,
        group-separator={,}
    }
    \caption{Performance comparison of different FDTD simulators for a simulation volume of $6\mu m \times 4 \mu m \times 1.5 \mu m$. Open-source software packages are marked in green.}
    \begin{tabular}{
        >{\columncolor{lightgray}}S
        >{\columncolor{lightgray}}S[table-format=3.1e1]
        >{\columncolor{lightgray}}S[table-format=5.0]
        c
        >{\columncolor{lightlightgreen}}c
        >{\columncolor{lightlightgreen}}c
        c
        >{\columncolor{lightlightgreen}}c
    }
        \toprule
        \rowcolor{lightgray}
        {\textbf{Resolution} \textbf{(nm)}} & 
        {\textbf{Cells}} & 
        {\textbf{Steps}} & 
        \textbf{OmniSim} & 
        \textbf{Ceviche} & 
        \textbf{Meep} & 
        \textbf{Tidy3D} & 
        \textbf{FDTDX}\\
        \midrule
        25.0 & 2.3e6 & 4196 & 7\,min\,5\,s & 8\,min\,30\,s & 32\,s & 1.6\,s & 4.7\,s \\
        20.0 & 4.5e6 & 5245 & 16\,min\,17\,s & 1\,h\,33\,min & 1\,min\,25\,s & 3.7\,s & 9.8\,s \\
        10.0 & 36.0e6 & 10490 & $\times$ & 23\,h\,49\,min & 21\,min\,20\,s & 1\,min\,47\,s & 1\,min\,50\,s \\
        5.0 & 288.0e6 & 20980 & $\times$ & 180\,h\,3\,min & 4\,h\,36\,min & 3\,min\,35\,s & 26\,min\,8\,s \\
        2.5 & 2.3e9 & 41960 & $\times$ & $\times$ & $\times$ & 14\,min\,26\,s & 7\,h\,5\,min \\
        \bottomrule
    \end{tabular}
    \label{tab:performance}
\end{table}

In \cref{tab:performance}, the runtime of different simulation frameworks is displayed for this setup.
This comparison is highly dependent on the hardware used and can change drastically.
We tried to make this comparison as fair as possible by using the strongest hardware available in our research group, which is compatible with the respective frameworks.
The open-source packages Ceviche and Meep were run on a CPU-Cluster with 16 cores (AMD EPYC, 4137MHz) and 256GB RAM.
For the highest resolution, both frameworks reached the memory limitations of this system.
OmniSim was run on a Windows consumer computer (Intel i7-8550U CPU/1.80 GHz, 16 GB RAM) due to licensing.
Resulting from the low memory, OmniSim was only able to run the simulations up to a resolution of $20nm$.
All three frameworks would be able to simulate finer resolutions on stronger hardware.
In contrast to the previous frameworks, Tidy3D uses a web-based service, where users can submit their simulation jobs online.
These jobs run on undisclosed special-purpose hardware, making comparisons difficult.
Additionally, runtimes may vary depending on the current demand.
For example, for the resolution of $10nm$, we observed a runtime between 1 minute 30 seconds and 2 minutes.
The values reported for Tidy3D in \cref{tab:performance} are average values for 3 runs at different days and times of day for resolutions $25$, $20$, and $10nm$.
For $5nm$, we performed only a single run for cost reasons.
For the run at $2.5nm$, we generously received funding from the Tidy3D developers themselves as a standard license is limited to 800 million grid points.
The pay-per-simulation model of Tidy3D would result in high costs for research projects with a large simulation demand.
This is exacerbated for gradient-based optimization which require many forward and backward simulations.
In contrast, our open-source FDTDX software can be run on in-house GPU, university compute clusters, or compute services like AWS if necessary.
However, using compute services will also result in costs which depend on the hardware used.
Since a cost comparison is difficult and highly dependent on the service provider, we do not consider it here.
We ran the FDTDX simulation on a single Nvidia H100 GPU for all resolutions up to $5nm$.
The simulation with $2.5nm$ resolution was run on four Nvidia H100 GPUs due to larger memory requirements.
The results of the speed comparison show that FDTDX outperforms all other open-source simulation software by a large margin.
At 288 million grid cells, it achieves a speedup of about 10x compared to Meep and 415x compared to Ceviche.
For 36 million grid cells, the speed of FDTDX is close to or even exceeds the speed of Tidy3D in some runs.
Only in very large simulations is FDTDX significantly slower than Tidy3D.
However, without calculating exact values, it is reasonable to assume that for large simulations, the price difference increases with the speed difference.
We believe that there is a place for both open-source and commercial software as open-source software is usually more cost-effective and gives researchers more freedom to implement their own algorithms.
In contrast, commercial software is often faster, but with the development of FDTDX, we make a significant step towards decreasing this gap.


In this speed comparison, we did not include any setup times.
For Tidy3D, we excluded the time required to send and setup the simulation and receive the results.
Additionally, we did not include waiting time in queue if demand is high.
For both Tidy3D and FDTDX, we did not include the time required to compile and optimize the computational graph.
This compilation step took between a single second for 2.3 million grid cells and two minutes for 2.3 billion grid cells.

\subsection{Silicon Waveguide Bend}
\begin{figure}[t]
    \centering
    \includegraphics[width=0.9\textwidth]{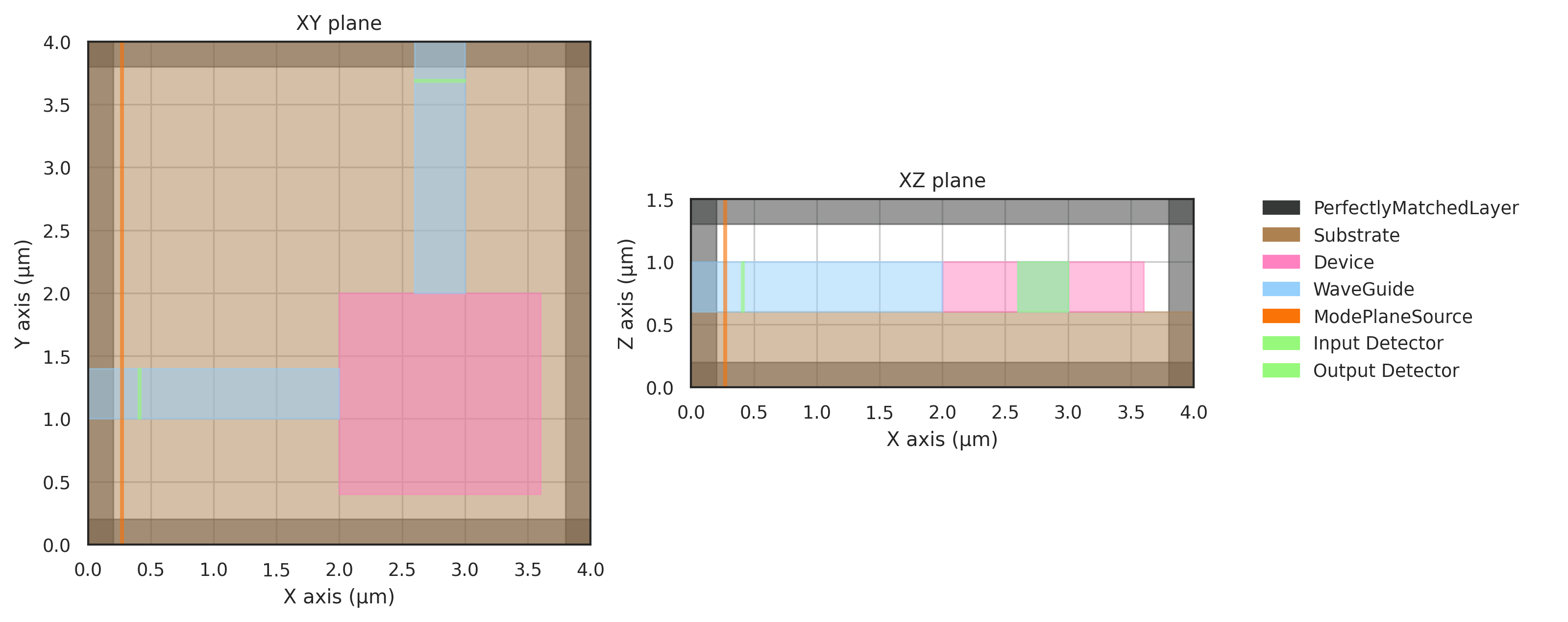}
    \caption{Simulation setup for a waveguide bend. The substrate consistis of silica with a refractive index of $1.5$, the waveguides of silicon with refractive index of $3.5$. Goal is to find a design of the device which redirects the light with as little loss as possible. The source has a wavelength of $1550nm$.}
    \label{fig:ceviche_corner_setup}
\end{figure}

In silicon photonics, routing of light is an important prerequisite for designing intricate photonic integrated circuits.
Ref.~\citenum{foundry_constraints} introduces several components to adress this problem.
Following their work, we optimize a waveguide bend that connects two waveguides oriented at a 90-degree angle.
The device has a small spatial footprint of $1.6 \mu m \times 1.6 \mu m$.
In \cref{fig:ceviche_corner_setup}, the setup for the simulation is visualized using the tools of our FDTDX-framework.
A mode source induces the first-order mode in the waveguide.
To calculate the efficiency of the device, the Poynting flux is measured at the start of the input waveguide and the end of the output waveguide.
The goal is to find a design of the device that redirects the light from the input to the output waveguide with as little loss as possible.
Discretizing the simulation volume of size $4 \mu m \times 4 \mu m \times 1.5 \mu m$ with a resolution of $20nm$ results in a total of 3 million grid cells.
The simulation was run for 200 femtoseconds, which are 5245 discrete time steps.

We optimized the design using our memory efficient automatic gradient computation with inverse time stepping.
This resulted in memory requirements for forward and backward simulation of about 20GB, making this optimization feasible on consumer graphics cards.
We chose to run the optimizations on a single NVIDIA H100 GPU.
A single gradient computation step using our direct differentiation approach took 100 seconds, resulting in a total runtime of 7.5 hours.
We followed the approach of Ref.~\citenum{foundry_constraints} to optimize a design adhering to minimal feature constraints.
For this optimization, we used a minimum feature constraint with respect to a circular brush of $100nm$.
The method using constraint enforcement at every step of the optimization process is compared to a two-stage process that only enforces constraints at the end of the optimization.
Specifically, we first optimized a design without enforcing any constraints other than the quantization to silicon or air.
After 250 optimization iterations, we started to constrain the device for minimum feature size and continued with 20 iterations of finetuning.
The purpose of the finetuning phase is finding minor improvements within the constrained design space.
This addresses some of the losses that arise due to the abrupt change in the valid design space.

\begin{figure}[tb]
    \centering
    \subfloat[]{\includegraphics[width=0.4\textwidth]{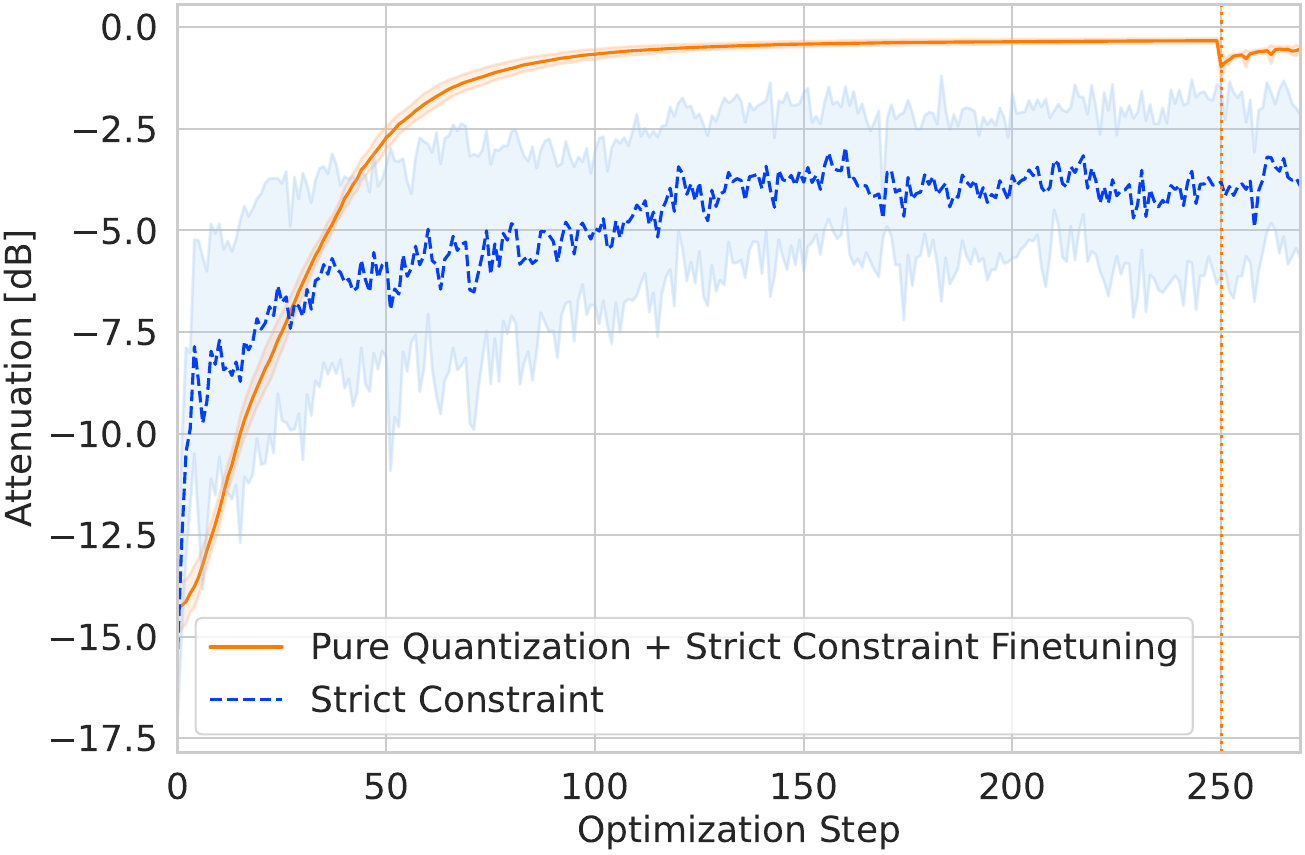}}
    \hspace{0.01\textwidth}
    \subfloat[]{\includegraphics[height=0.27\textwidth]{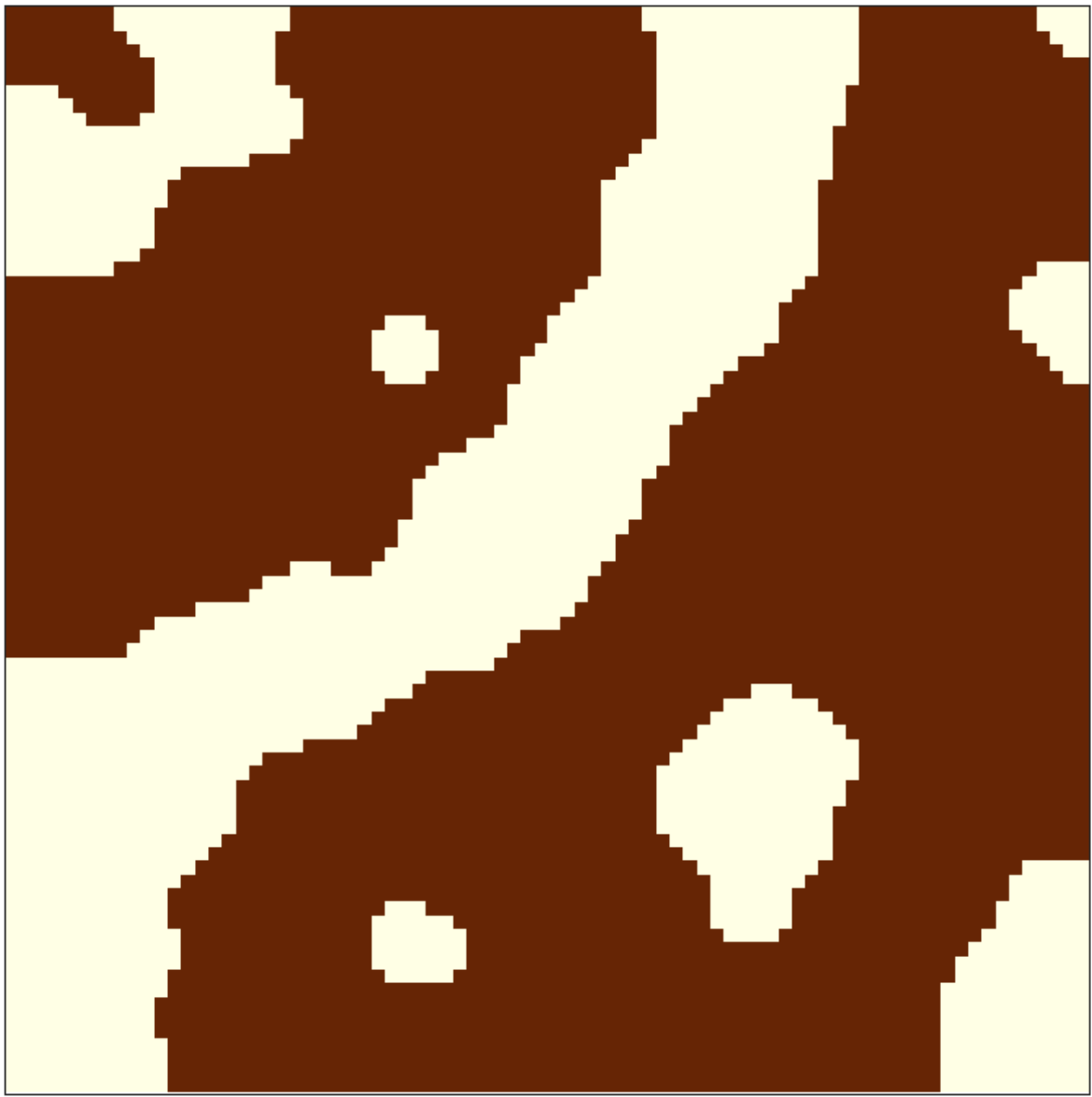}}
    \hspace{0.01\textwidth}
    \subfloat[]{\includegraphics[height=0.27\textwidth]{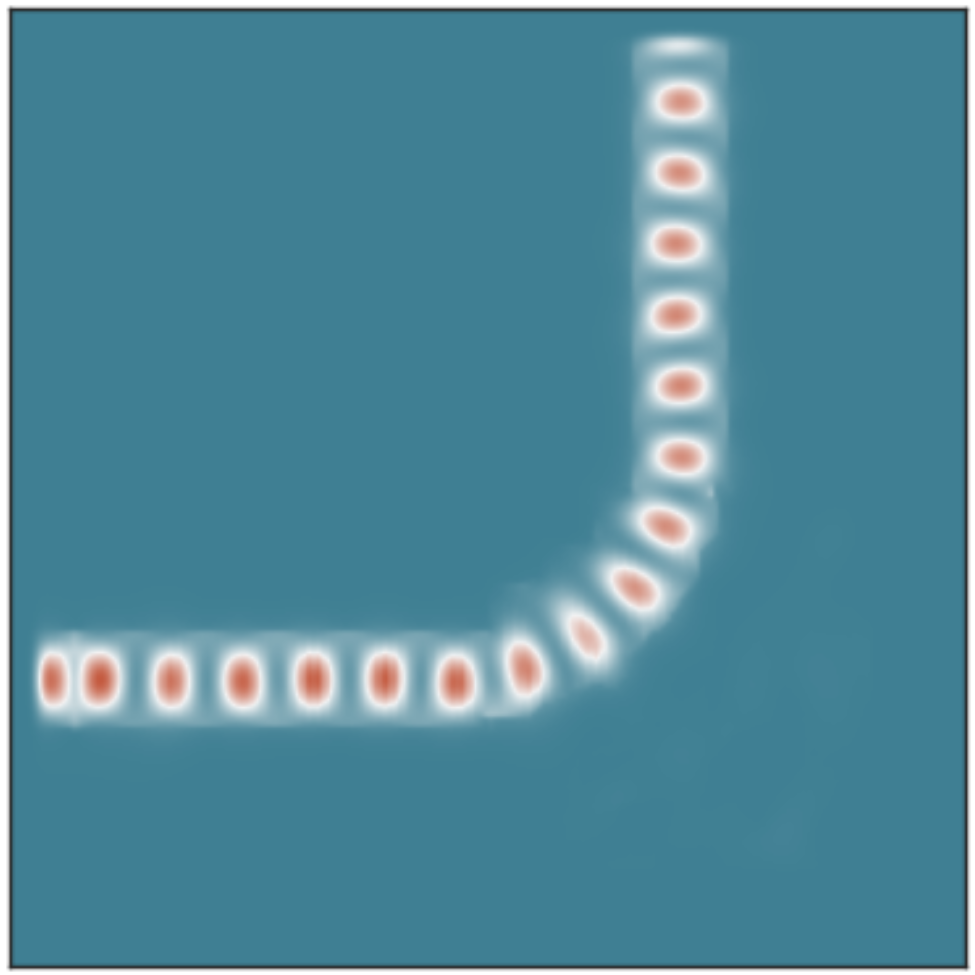}}
    \caption{Optimization of a waveguide bend. 
    In (a), optimization results are presented using either strict constraints (blue) or a two-stage approach: an initial unconstrained optimization followed by finetuning within the constrained design space after 250 steps (orange).
    Mean and standard deviation are calculated over five random initializations. 
    The best performing device is visualized in (b), where silicon is visualized in a brown color.
    In (c), the energy distribution using this design is shown.}
    \label{fig:ceviche_corner_results}
\end{figure}

The results are displayed in \cref{fig:ceviche_corner_results}.
We found that optimizing the waveguide bend using the strict foundry constraints enforced at every optimization step leads to a high variance in the results.
Some optimization runs resulted in a high-performance device, while others did not.
Using two-stage optimization yielded more reliable results because the unconstrained devices were already close to adhering to the constraints.
Consequently, only a few decibels of loss incurred when strictly constraining the devices.
The final device with the best performance achieved an attenuation of -0.36dB, or equivalently an efficiency of $92\%$.
Further research is necessary to prove whether the finetuning approach is superior to strict constraints in other simulation scenes as well.

\subsection{Polymer Waveguide Stitching Device}
As demonstrated in the previous experiment, silicon structures can exhibit considerable functionality.
However, silicon devices are suboptimal for prototyping because fabrication is time-consuming and expensive.
In contrast, Two-Photon-Polymerization (2PP) offers low-cost prototyping using polymers with the additional benefit of fully three-dimensional photonics.
Although the possibility of three-dimensional design offers more degrees of freedom and enables devices with new functionality, it introduces new design challenges as well.
In addition to the minimum feature constraint of two-dimensional designs, three-dimensional designs need to adhere to two other constraints.
Firstly, the design needs to be constrained such that all material is connected to the ground.
Additionally, due to the fabrication process, a three-dimensional design for 2PP cannot have any fully enclosed air cavities.
Such cavities would trap non-polymerized monomer, which would weaken the structural integrity.
Ref.~\citenum{schubert2024quantized} presented algorithms for incorporating these constraints into the optimization process.

\begin{figure}[htbp]
    \centering
    \includegraphics[width=0.9\textwidth]{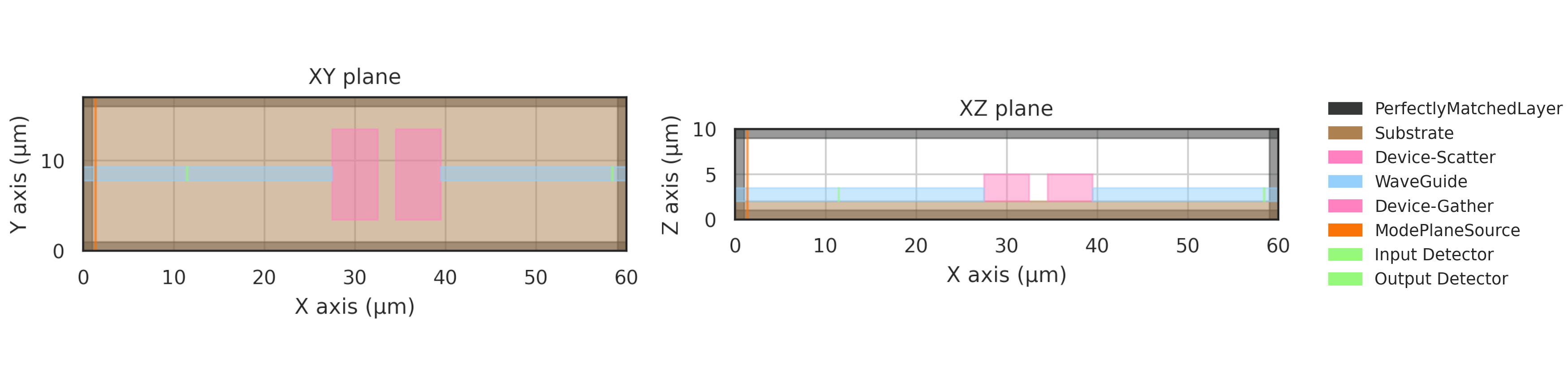}
    \caption{
    Simulation scene for a stitching element robust to random translations.
    The mode source inserts light into the input waveguide.
    The efficiency of the stitching element is measured as the fraction of Poynting flux measured at the input and output detector.
    }
    \label{fig:stitching_setup}
\end{figure}

To demonstrate the capabilities of the FDTDX framework in three-dimensional design optimization, we addressed the fabrication of long waveguides using 2PP.
The two-photon polymerization process, implemented through systems such as Nanoscribe, employs objectives of varying magnifications.
Objectives with higher magnification can write finer structures but have a smaller field of view and, therefore, smaller effective printing radius.
Fabrication of larger structures requires the stitching of multiple printing areas, which introduces artifacts and printing errors \cite{micro1020013}. 
These imperfections particularly affect the production of long waveguides, where misalignments of just a few micrometers can significantly degrade performance.
We optimized a stitching device that is resistant to random translational offsets.
For this experiment, we assumed a maximum offset of $2\mu m$ in the x and y directions.

In \cref{fig:stitching_setup}, the simulation setup for the stitching device is shown.
The design consists of two parts.
A scattering device is placed at the end of the input waveguide, which conceptually collimates the light from the waveguide into a planar free-space wave.
At the beginning of the output waveguide, a gathering device redirects the collimated light back into the polymer waveguide.
The gathering device and the output waveguide are randomly moved by up to $2\mu m$ in the x- and y-directions to simulate stitching errors.

\begin{figure}[bt]
    \centering
    \subfloat[]{\includegraphics[width=0.31\textwidth]{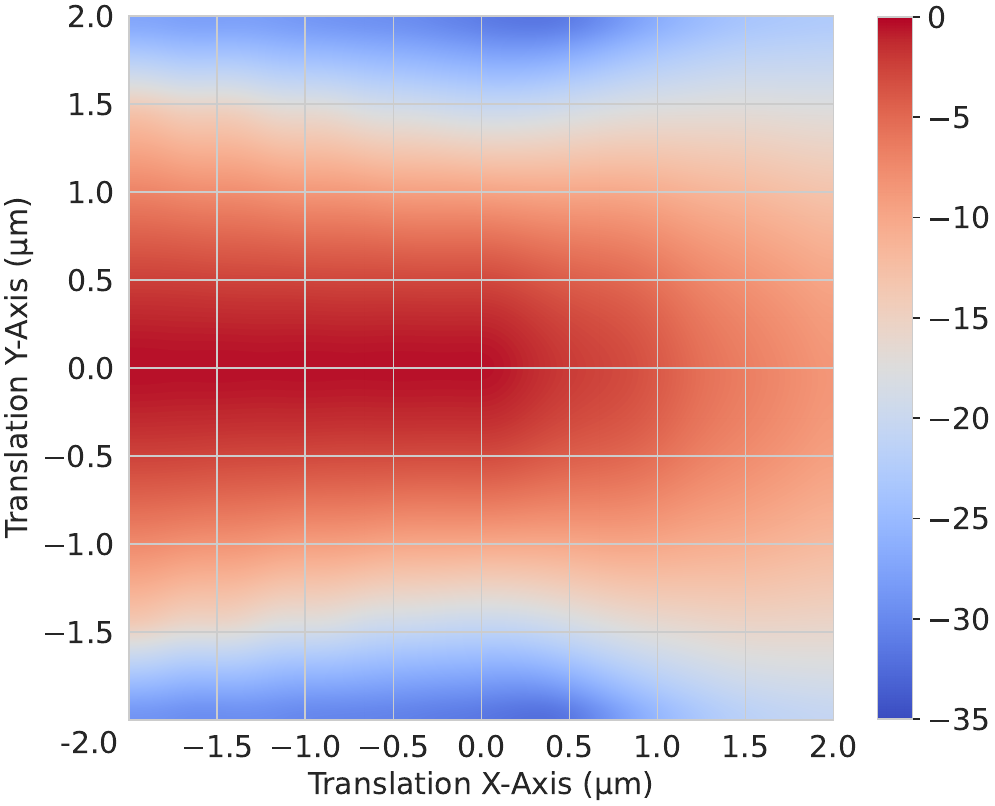}}
    \hspace{0.01\textwidth}
    \subfloat[]{\includegraphics[width=0.31\textwidth]{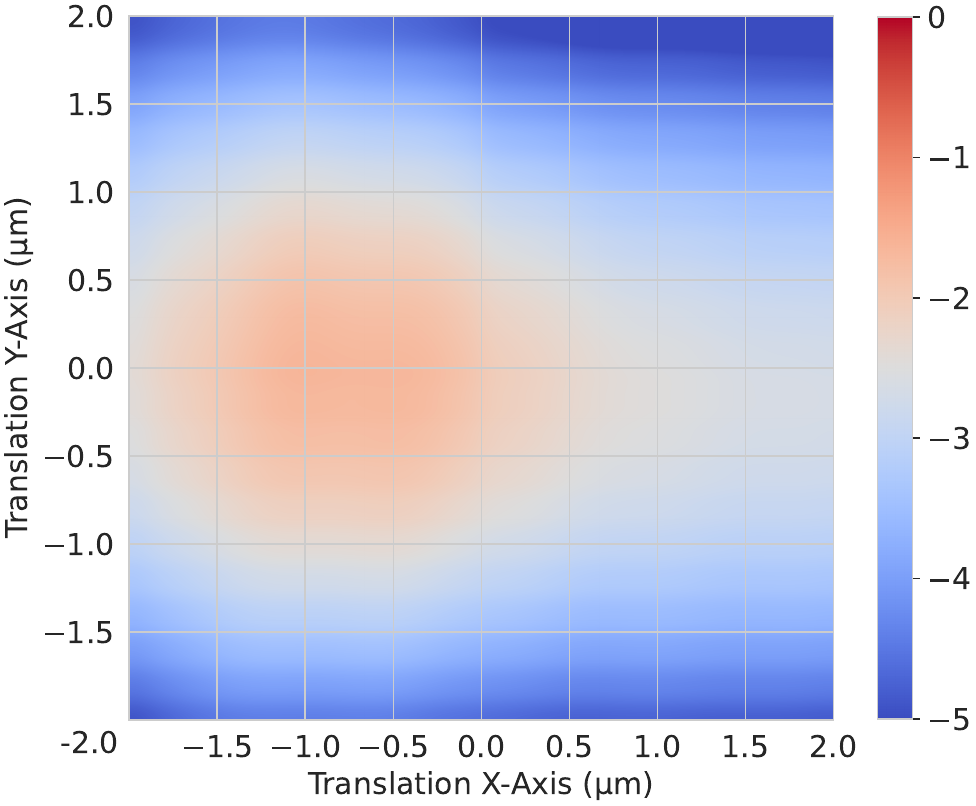}}
    \hspace{0.01\textwidth}
    \subfloat[]{\includegraphics[width=0.31\textwidth]{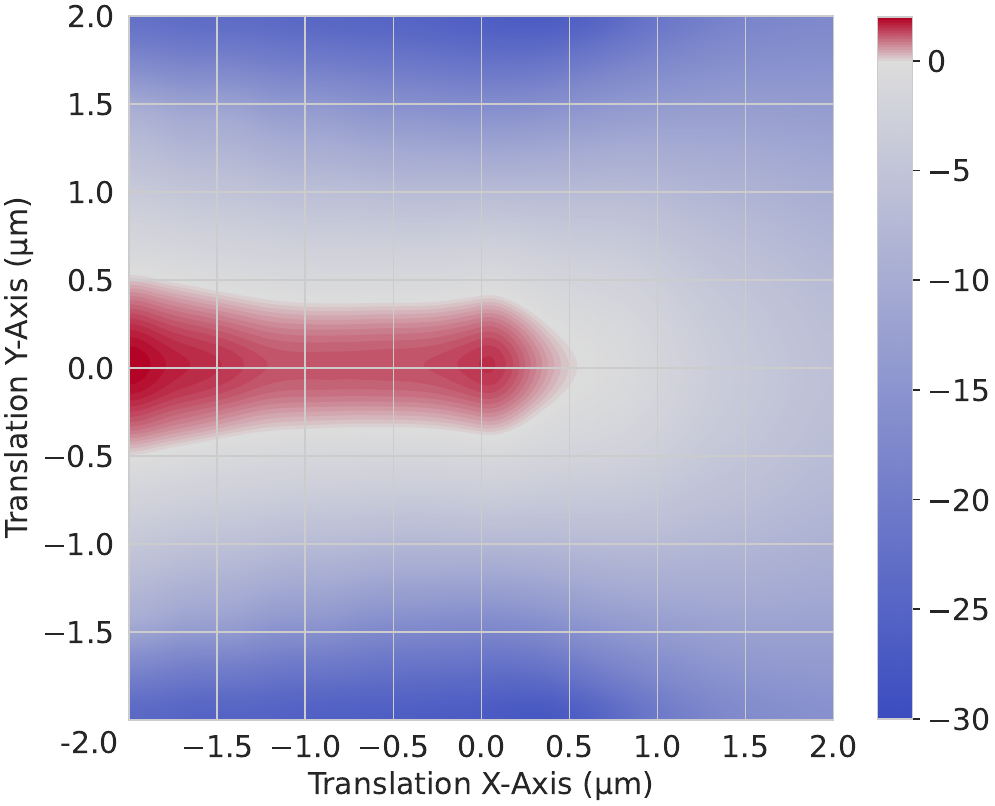}}
    \caption{
    Analysis of waveguide stitching under random translation in the x- and y-axis.
    In (a), the coupling attenuation of a standard ridge waveguide is analyzed.
    In (b), the attenuation is measured for our optimized design.
    We display the average performance over five random initializations.
    The difference between the standard waveguide and our design is displayed in (c).
    }
    \label{fig:waveguide_stitching_analysis}
\end{figure}

\begin{figure}[tb]
    \centering
    \subfloat[]{\includegraphics[width=0.58\textwidth]{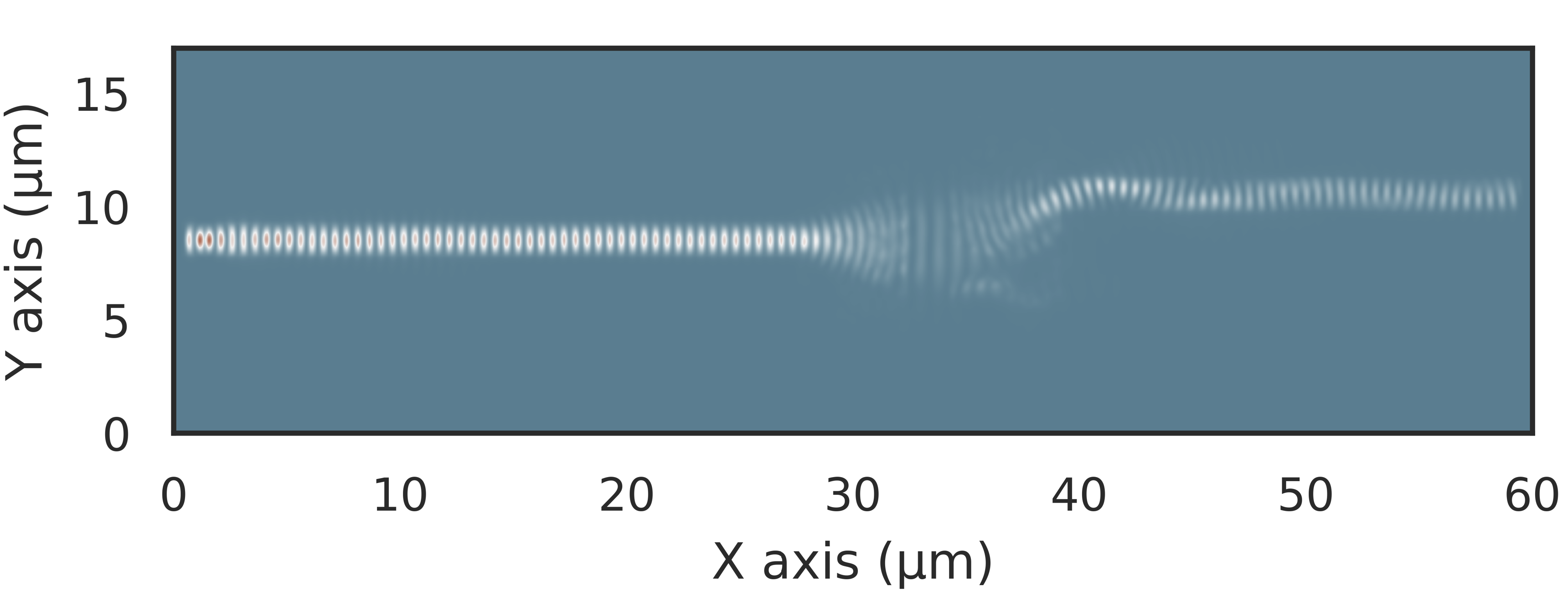}}
    \hspace{0.01\textwidth}
    \subfloat[]{\includegraphics[width=0.4\textwidth]{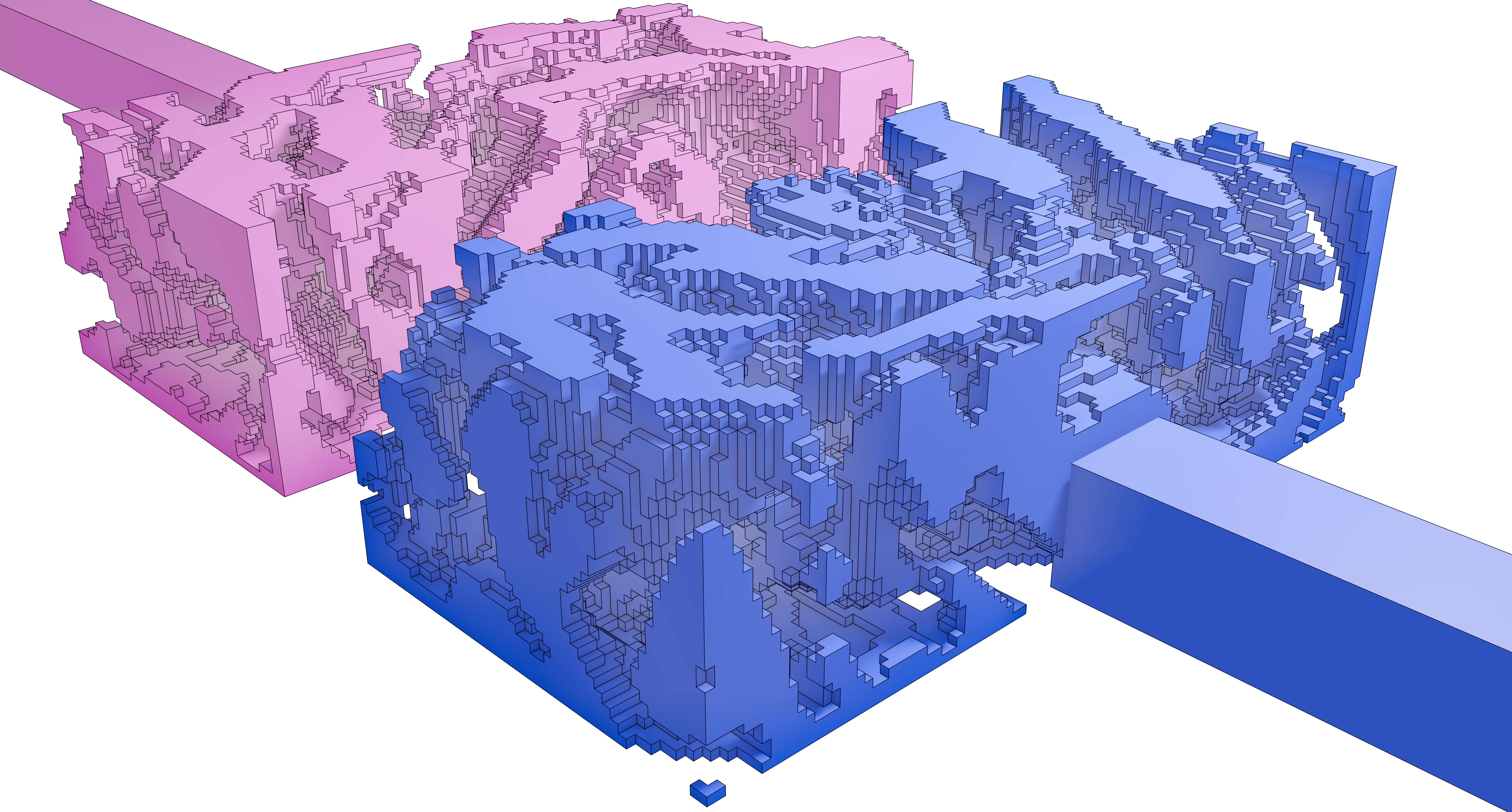}}
    \caption{
    (a) Energy distribution of the optimized design and a large offset of $2 \mu m$ in y-direction.
    Input light is collimated from the waveguide into free space and redirected into the translated waveguide.
    In (b), a rendering of the optimized waveguide stitching device is shown. 
    The left part (pink) is the input waveguide and scattering device that collimates light into free space.
    The right part (blue) redirects the collimated light back into the waveguide.
    }
    \label{fig:stitching_rendering}
\end{figure}

For optimization, we used a slightly more compact simulation scene of size $40\mu m \times 17\mu m \times 10\mu m$ and resolution of $100nm$, which results in a total of 6.8 million grid cells.
Each gradient step required 28 seconds on a single NVIDIA H100.
We optimized the design for 500 gradient steps, resulting in a total runtime of 3 hours and 53 minutes.
The simulation was run for 300 femtoseconds at a courant factor of 0.99, resulting in 1574 discrete time steps.
The memory requirements for this optimization were about 15GB, again making this optimization feasible on consumer graphics cards.
We used strict constraints at every time step, because in three-dimensional designs it is not feasible to combine unconstrained optimization with finetuning due to the additional constraints.
In \cref{fig:waveguide_stitching_analysis}, we compare the performance of our optimized designs to that of a standard waveguide.
The standard waveguide has an attenuation close to 0dB when the random translation is low, but the attenuation drops to -35dB for an x-translation of $2\mu m$.
In contrast, our optimized design exhibits an attenuation between -1dB and -5dB.
The design does not achieve a perfect transmission when no random translation is present as some light is lost in the collimation process.
Therefore, it is more efficient to use a standard waveguide when errors are small and to use our design when errors are large.
For random translation in the x-direction, our device is only more efficient if an air gap between emerges in the standard waveguide due to a translation to the right.
However, when a normal waveguide is used, this can be avoided simply by moving the whole waveguide to the left.
But, our device is much more efficient than the standard waveguide when the random translations are larger than $0.5\mu m$ in the y-direction.
In \cref{fig:stitching_rendering}, a rendering of the optimized design is shown.

In future work, it would be interesting to optimize a design that is also robust against random offsets in the z-axis.
These offsets can occur due to misalignments when moving to a different printing field, but also due to skewered substrate.
Moreover, the coupling efficiency of our device could be improved if there is no air gap between the left and right parts of the design.
This would require simulating an overlap of the two devices if the random translation moves the parts closer together, which we leave to future work.

\section{OPEN-SOURCE FDTD IMPLEMENTATIONS}

There exist many different software frameworks for near-field electromagnetic simulation and optimization using the FDTD method.
The popular open-source implementation Meep (MIT Electromagnetic Equation Propagation) \cite{meep} was developed in 2006 and is still widely used today.
Other open-source frameworks for FDTD simulation are OpenEMS \cite{openEMS} and EMopt \cite{emopt}, which are also implemented in C++.
Ceviche \cite{ceviche} is a framework based on the python numpy library \cite{numpy} and supports automatic differentiation.
Further frameworks, which are no longer supported, include Semba-FDTD \cite{semba_fdtd}, which is written in Fortran, and Spins-B \cite{spins} written in C++.
However, all of the previously mentioned frameworks are not compatible with GPUs.
There exist a few open-source frameworks that support execution on GPU, for example the FDTD framework by Ref.~\citenum{fdtd_laporte}, which has an optional PyTorch backend \cite{pytorch}.
Luminescent \cite{luminescent} and Khronos \cite{khronos} are two FDTD frameworks written in the Julia programming language with GPU support.
However, all of these frameworks either only support a single GPU or do not support automatic differentiation.
Another framework, which is no longer actively maintained, but supports GPU is GSvit \cite{gsvit}.
FDTD-Z, which is also not actively maintained, takes the efficient approach of implementing custom cuda kernels for execution on graphics processing units \cite{fdtdz}.
Unfortunately, the feature set of this framework is very limited andit only supports a single graphics card.
In future work, we plan to integrate such custom kernels into our framework to increase the execution speed even further.
In addition to all the mentioned open-source frameworks, there also exist a variety of commercial software for FDTD simulations.
We do not go into more detail here as it is difficult to compare the feature set of commercial software without obtaining expensive licensing.
Additionally, we believe open-source implementations can best advance research in the area of photonics.
In addition to FDTD, there also exist numerous implementations of the finite-difference frequency-domain (FDFD), which we do not cover here.

\section{CONCLUSION}

We presented FDTDX, an open-source framework for large-scale electromagnetic simulations and inverse design that addresses several key limitations of existing FDTD software. 
Through our implementation of memory-efficient automatic differentiation based on time-reversibility, FDTDX enables gradient-based optimization of complex 3D nanostructures that would be infeasible to design manually. 
The key innovations include a flexible reverse-mode automatic differentiation implementation that drastically reduces memory requirements by leveraging the time-reversibility of Maxwell's equations.
Additionally, a novel relational object API simplifies the specification of complex 3D simulation scenes through intuitive positioning and sizing constraints.
Our framework scales from single to multiple GPUs, enabling simulations with billions of grid cells.
It enables efficient optimizations for both two- and three-dimensional photonic design optimization, as demonstrated through a silicon waveguide bend and polymer waveguide stitching device.
Our experimental results validate both the accuracy and performance advantages of FDTDX.
For future work, we plan to integrate custom CUDA kernels to further improve computational performance.
Additionally, we plan to expand the feature set of our framework to match and possibly even exceed the feature set of established tools like Meep.
Lastly, we plan to develop a graphical user interface to make electromagnetic simulation and optimization accessible to users without programming experience.
By releasing FDTDX as open-source software, we aim to democratize access to advanced electromagnetic design capabilities and accelerate innovation in nanophotonics. 
The combination of performance, ease of use, and flexibility makes it a valuable tool for researchers in the field, from photonic integrated circuits to metamaterial design.

\appendix    

\section{Validation comparison to meep}

\begin{figure}[htbp]
    \centering
    \includegraphics[width=\textwidth]{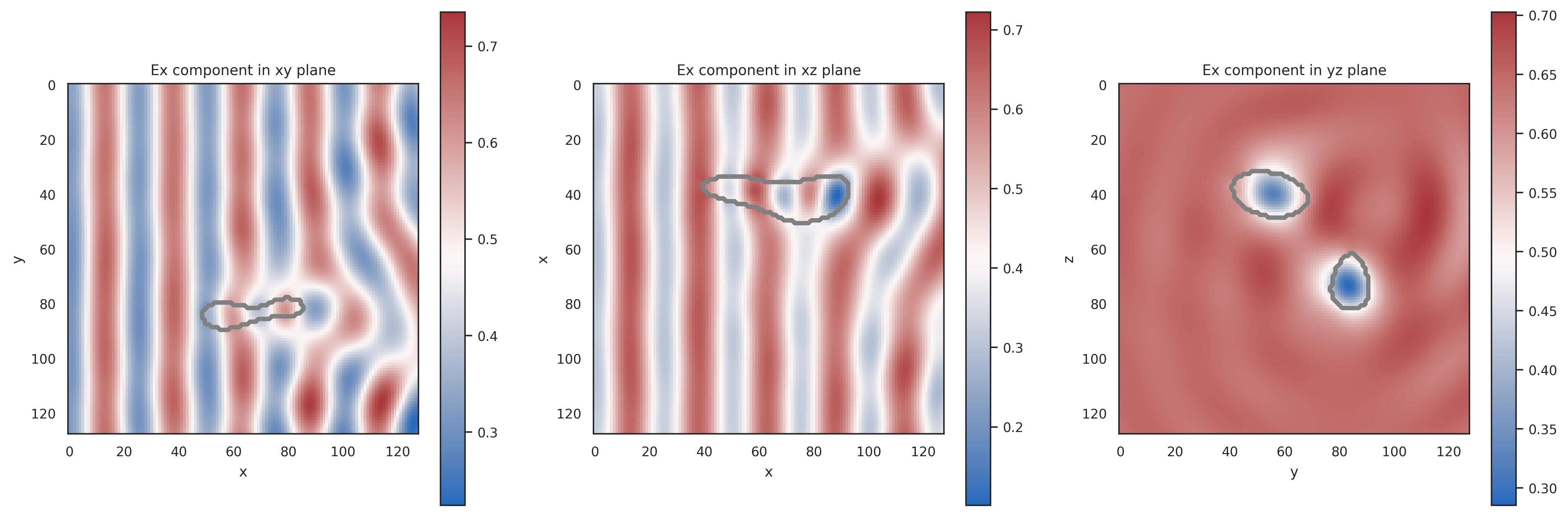}
    \includegraphics[width=\textwidth]{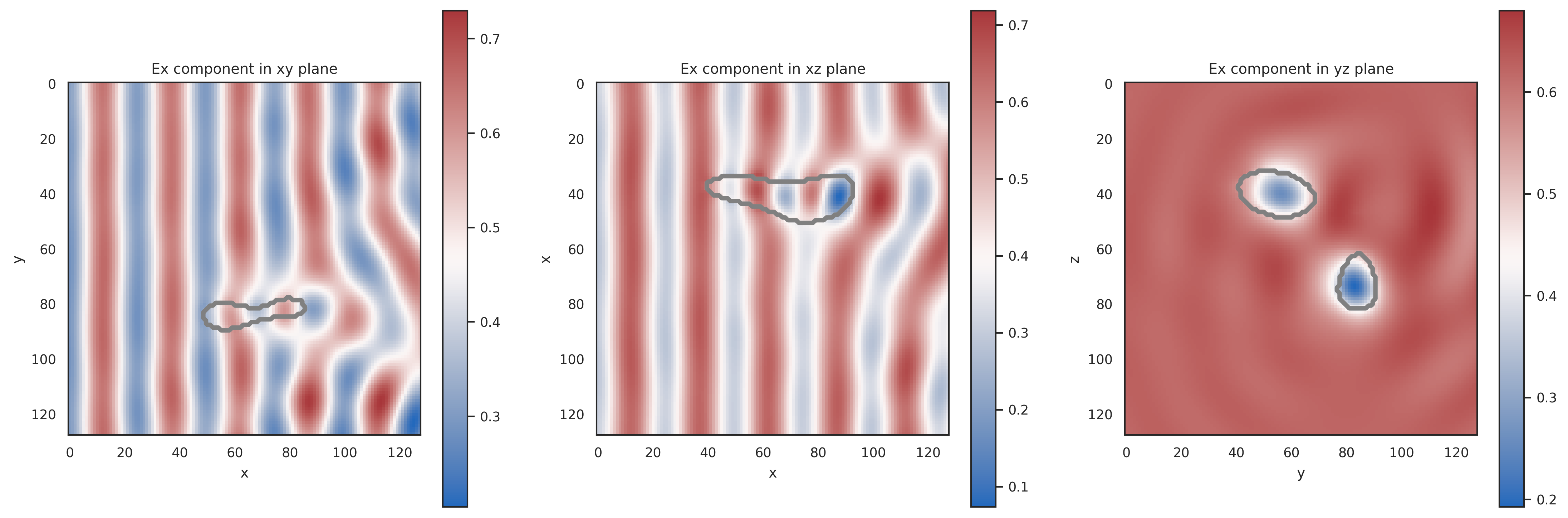}
    \includegraphics[width=\textwidth]{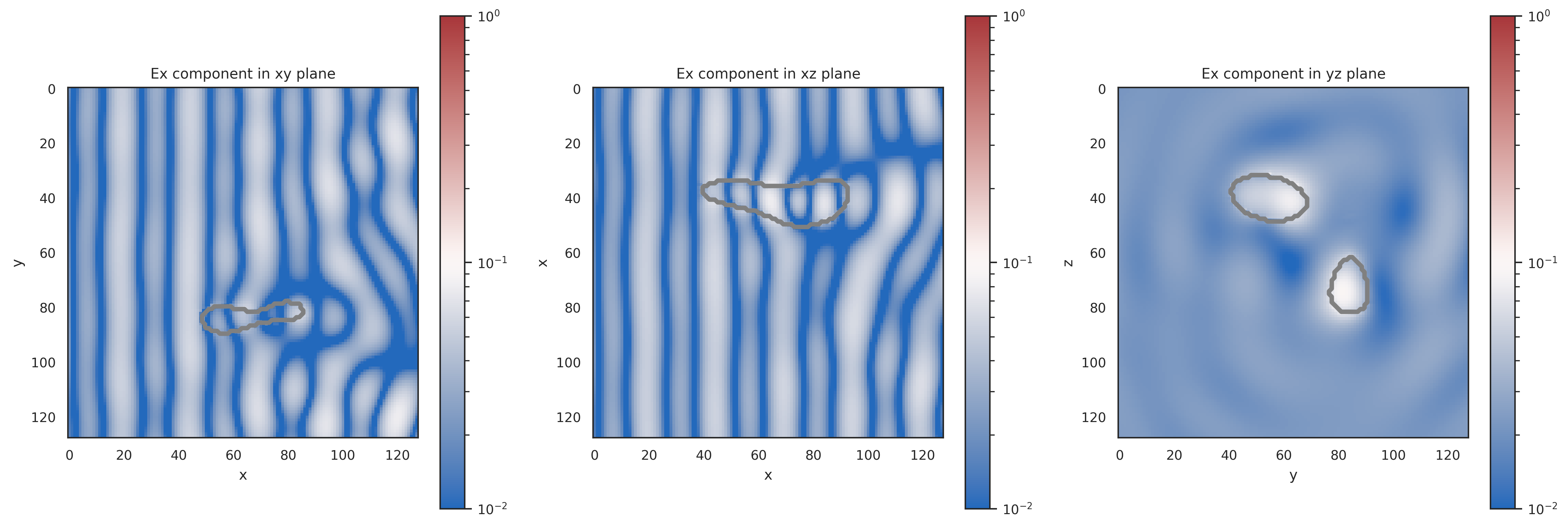}
    \caption{
    Distribution of the electric field component in x-direction measured in the XY, XZ and YZ plane at the center of the simulation volume.
    A random scattering device was simulated.
    The top row shows results of our simulations using FDTDX, the middle row the results generated with Meep \cite{meep}.
    The bottom row shows the normalized error between the two simulations.
    }
    \label{fig:validation}
\end{figure}

To validate the results of our FDTDX framework, we reproduce simulations performed in Meep \cite{meep}.
To this end, we simulated light scattering on randomly generated objects.
Originally, these simulations were used in a dataset of field distributions to train a neural operator \cite{neural_operator}.
For these simulations, a volume of $6.12\mu m^3$ was simulated.
The random scattering object with refractive index $1.5$ was placed in the center of the simulation volume.
At the bottom, a planar wave with wavelength $1\mu m$ induced light for 10 periods.
The field distributions at the three centered planes of the simulation were recorded until all of the energy dissipated.
We recreated these simulations with our framework and compared our results with the results reported in the dataset.
In \cref{fig:validation}, an example for a single random scattering object is shown.

\acknowledgments 
 
This work was supported by the Federal Ministry of Education and Research (BMBF), Germany under the AI service center KISSKI (grant no. 01IS22093C), the Lower Saxony Ministry of Science and Culture (MWK) through the zukunft.niedersachsen program of the Volkswagen Foundation and the Deutsche Forschungsgemeinschaft (DFG) under Germany’s Excellence Strategy within the Cluster of Excellence PhoenixD (EXC 2122) and  (RO 2497/17-1). Additionally, this was funded by the Deutsche Forschungsgemeinschaft (DFG, German
Research Foundation) - 444745111 and 517733257.

\bibliography{references} 
\bibliographystyle{spiebib} 

\end{document}